\documentclass[%
 reprint,
nofootinbib,
 amsmath,amssymb,
 aps,
floatfix,
]{revtex4-2}

\usepackage{macros} %for bib
\usepackage{graphicx}% Include figure files
\usepackage[caption=false]{subfig}
\usepackage{dcolumn}% Align table columns on decimal point
\usepackage{bm}% bold math
\usepackage{multirow} %multirow for tables
\usepackage{float} % for positioning table
\restylefloat{table} % for positioning table
\usepackage{listings}
\usepackage{soul}
\usepackage{xcolor} % Colours
\definecolor{blue}{rgb}{0.36, 0.54, 0.85}
\definecolor{amaranth}{rgb}{0.9, 0.17, 0.31}
\definecolor{pink}{rgb}{0.87, 0.56, 0.81}
\definecolor{ao}{rgb}{0.0, 0.5, 0.0}
\definecolor{maroon}{rgb}{0.76, 0.13, 0.28}
\definecolor{cardinal}{rgb}{0.77, 0.12, 0.23}
\definecolor{lightcardinal}{rgb}{0.97, 0.42, 0.53}
\definecolor{frenchlila}{rgb}{0.53, 0.38, 0.56}
\definecolor{pink}{rgb}{1.0, 1.0, 0.87}
\definecolor{lightseagreen}{rgb}{0.45, 0.85, 0.58}
\definecolor{gray}{rgb}{0.9, 0.9, 0.9}
\definecolor{lightblue}{rgb}{0.66, 0.84, 0.96}

\usepackage[pagebackref=false,
			hyperindex=true,
			citecolor=blue,
			colorlinks=true,
			linkcolor=blue,
			urlcolor=blue]{hyperref}

\newcommand{\param}{\ensuremath{\vec{\lambda}} }

\newcommand{\mrm}[1]{\mathrm{#1}}

\begin{document}

\preprint{APS/123-QED}

\title{Exploring the Potential for Detecting Rotational Instabilities in Binary Neutron Star Merger Remnants with Gravitational Wave Detectors}% Force line breaks with \\

\author{Argyro Sasli$^{1}$, Nikolaos Karnesis$^{1}$ and Nikolaos Stergioulas$^{1}$\\
% List of institutions
$^{1}$ Department of Physics, Aristotle University of Thessaloniki, Thessaloniki 54124, Greece}%
% \affiliation{%
%  Authors' institution and/or address\\
%  This line break forced with \textbackslash\textbackslash
% }%

\date{\today}% It is always \today, today,

\begin{abstract}
We explore the potential for detecting rotational instabilities in the post-merger phase of binary neutron star mergers using different network configurations of upgraded and next-generation gravitational wave detectors. Our study employs numerically generated post-merger waveforms, which reveal the re-excitation of the $l=m=2$ $f$-mode at a time of $O(10{\rm})$ms after merger. 
We evaluate the detectability of these signals by injecting them into colored Gaussian noise and performing a reconstruction as a sum of wavelets using Bayesian inference. Computing the overlap between the reconstructed and injected signal, restricted to the instability part of the post-merger phase, we find that
one could infer the presence of rotational instabilities with a network of planned 3rd-generation detectors, depending on the total mass and distance to the source. For a recently suggested high-frequency detector design, we find that the instability part would be detectable even at 200 Mpc, significantly increasing the anticipated detection rate. For a network consisting of the existing HLV detectors, but upgraded to twice the A+ sensitivity, we confirm that the peak frequency of the whole post-merger gravitational-wave emission could be detectable with a network signal-to-noise ratio of 8 at a distance of 40Mpc.

\end{abstract}

\maketitle

%\tableofcontents

\section{Introduction\label{sec:introduction}}

The detection of GW150914 \cite{firstGW}, which was the first observed gravitational wave (GW) event, initiated the beginning of GW astronomy. With the completion of three observing runs (O1, O2 and O3) by the ground-based detectors advanced LIGO \cite{LIGO} and advanced Virgo \cite{Virgo}, recently joined by KAGRA \cite{KAGRA,KAGRA1} more than 90 GW events are included in the Gravitational Wave Transient Catalog (GWTC) \citep{GWTC,GWTC1,GWTC-2,GWTC-4,GWTC-3,GWTC-5} (see \cite{Nitz_2023} for additional candidates),  two of which, GW170817 \cite{GW170817} and GW190425 \cite{GW170817, GW190425}, were identified as Binary Neutron Star (BNS) mergers. This has opened new avenues for probing the physics at the extreme densities encountered at the center of compact stars (for comprehensive reviews, see, e.g.~\cite{Baiotti_review,Friedman_Stergioulas_2020,2021GReGr..53...59S, Baiotti_2022,Arimoto:2021cwc} and references therein). The detection of new BNS merger signals in the coming observational runs (such as the ongoing O4 run) is expected to further refine our understanding of compact stars and their Equation of State (EoS) (e.g., see~\cite{EoS_improve1, EoS_improve2, EoS_improve3, EoS_improve4, EoS_improve5, Bhaskar_2021, 2022PhRvL.128p1102B, PhysRevD.102.123023, PhysRevD.100.044047, Bauswein_Blacker_2020, Iacovelli:2023nbv, Kunert:2021hgm,Chatziioannou:2021tdi,Chatziioannou:2020msi,Landry:2020vaw,Chatziioannou:2019yko} for how we could benefit from further observations).

GW170817 was the first GW event that allowed us to constrain the EoS of neutron stars (NS) and rule out some very soft models of high-density matter \cite{GW170817, Bauswein_2017, BNS_prop1, BNS_prop2, BNS_prop3,2022PhRvD.105j3022K, Friedman_Stergioulas_2020, Kastaun:2021zyo,Kastaun:2019bxo,Chatziioannou:2020pqz}. Moreover, this event marked the beginning of the era of multi-messenger astronomy, since it was detected with both Gravitational and Electromagnetic (EM) radiation~\citep{GW170817_multi, Branchesi_GW17_multi}, spanning from radio, through optical and x-rays \cite{GW170817_xrays} to gamma rays \cite{GW170817_grays,2017ApJ...848L..14G}. More such multi-messenger events are expected to be observed in the future observing runs (e.g. \cite{EM_BNS_O4}) and provide us with the opportunity to combine different information for the same event\cite{Miller_2021,Raaijmakers_2020,PhysRevLett.120.261103,PhysRevD.101.123007,Traversi_2020,PhysRevD.103.103015,Dietrich_2020,Bulla:2022ppy,Breschi:2023ttc,Breschi:2021wzr}.

Despite the improvement in the sensitivity of current detectors to the O4 level, there are no high hopes of detecting the post-merger part of the GW waveform of BNS mergers during O4~\cite{Abadie_2010, PhysRevD.90.062004, Chatziioannou_2017,Torres_Rivas_2019, Fiona_2023}. However, such signal could be detected with future detector upgrades, beyond the fifth observing run (O5)
and more likely with the planned third-generation detectors \citep{BNS_3G_1,BNS_3G_2,BNS_detections,BNS_detection2,CE_BNS_2022, 2022arXiv220509979B}. The importance of such post-merger waveform signals acts as a motivation for dedicated observatories, such as the one suggested in~\cite{HF}, which we will henceforth call the High Frequency (HF) detector design.

The detection of GWs in the post-merger phase of BNS mergers will allow us to place constraints on the NS EoS in a density regime that cannot be probed directly by extracting the tidal deformability of NS in the inspiral phase. Moreover, the post-merger phase is rich in additional physics that could be probed (high temperature, shock waves, magneto-rotational instabilities, unstable oscillation modes, etc.).

Simulations of BNS mergers have shown~\cite{Tootle_2021, PhysRevD.71.084021, Kolsch:2021lub}, that the remnant can either promptly collapse to a black hole (BH), survive for a short duration (of order several or tens of milliseconds) before a delayed collapse takes place, or even avoid collapse all together and remain as stable NS, if the component masses are sufficiently small and the EoS sufficiently stiff. In the post-merger phase, the highly deformed remnant continues for some time to radiate GWs, if it is in a quasi-stable state or a stable state. There are different studies, such as Ref.~\cite{Agathos2020,2023CQGra..40v5008T}, that implement techniques to predict whether the merger will form a remnant or will promptly collapse into a black hole based on the gravitational-wave signal. 

The strongly differentially rotating remnant can avoid prompt collapse if the total mass of the binary is above a threshold mass. The remnant then becomes a transient Hypermassive NS (HMNS)~\cite{Baumgarte_2000, Hotokezaka_2013, PhysRevD.96.063011,Bauswein_2016, Kastaun:2014fna}.
The merger process excites linear nonaxisymmetric oscillation modes, nonlinear combination tones, and other transient effects in the remnant; see e.g. \citep{Stergioulas_2011, Hotokezaka_2013, Bauswein_2015, Takami_2015, Bauswein_2016, Bauswein_2019, Paschalidis, Clark_2016, PhysRevD.94.064011, De_Pietri_2018, Fields_2023,Kastaun:2010vw} and references therein. The main oscillation mode is the  $l=m=2$ $f$-mode, the frequency of which is typically denoted as $f_{\rm{peak}}$ in the literature on post-merger GW spectra\footnote{In some cases it is also denoted as $f_2$.}.

The excitation of these modes leaves a unique signature in the GW spectrum, typically between $\mathrm{2~kHz}$ and $\mathrm{4~kHz}$, which depends on the masses of the two binary components and the EoS, see, e.g. ~\citep{PhysRevLett.94.201101,PhysRevD.71.084021, PhysRevLett.99.121102, Bauswein_Janka, Bauswein_2012, Hotokezaka_2013,Takami_2014, Bauswein_2015,Takami_2015,Bauswein_2019,Friedman_Stergioulas_2020,Fields:2023bhs} and references therein. Thus, the oscillation modes are closely related to the nature of the resulting NS remnant, and by analyzing the information of the GW spectrum, one can infer various characteristics of the NSs, e.g. ~\cite{PhysRevLett.94.201101, 2023arXiv231010728T, 1992ApJ...401..226R,PhysRevD.90.062004,PhysRevD.93.124051, PhysRevD.93.044019} and \cite{Bauswein_2019} for a review and references therein.

For example, there are empirical relations that correlate $f_{\rm peak}$ with the radius of the non-rotating NS in the inspiral phase~\cite{Bauswein_2012, PhysRevD.100.104029, Hotokezaka_2013, Takami_2015, Bernuzzi_2015, Bose_2018, Bauswein_2019, PhysRevD.100.104029, Vretinaris_2020}. A recent study ~\cite{Criswell_2023} demonstrates how to employ such empirical relations to extract $f_{\rm peak}$, when combining information from several subthreshold events. The hierarchical Bayesian method proposed in~\cite{Criswell_2023} overcomes the limitations posed by a low signal-to-noise ratio (SNR) in retrieving $f_{\rm peak}$.

Apart from stable oscillations excited during merger, {\it rotational instabilities} could also develop in the remnant (see \citep{2013rrs..book.....F} and references therein). Of particular interest are the {\it dynamical shear instabilities}, often manifested as the {\it low$-|T/W|$ rotational instabilities}
~\cite{Centrella_2001,Passamonti_2020, Watts_Andersson_Jones} where $T$ is the rotational kinetic energy and $W$ the gravitational binding energy of the star.  Differentially rotating stars can develop such instabilities for relatively low values of the rotational parameter $\beta = |T/W|$,
\cite{Centrella_2001,Shibata_2002, Shibata_2004, Saijo}. In particular, one condition for the low$-|T/W|$ instability to set in, is that the pattern speed of the $l=m=2$ $f$-mode matches the rotational speed
of the star \cite{Watts_Andersson_Jones,Saijo_Yoshida, Corvino_2010, Passamonti_2015, Saijo_2016, Passamonti_2020}. 

Due to the instability's nature, depending on the rotational parameter, the impact of different rotation laws of HMNSs on the low$-|T/W|$ rotational instability was studied by~\cite{Passamonti_2020} and they attempted to give a qualitative explanation of its features observed in numerical simulation of either BNS remnants (e.g., the Ref.~\cite{DePietri_2020, Xie}) or rapidly rotating cold NSs (e.g., the Ref.~\cite{Corvino_2010}). It was found that in a differentially rotating BNS merger remnant, an oscillation mode could
co-rotate with the matter at two different points. The parameter $\beta$ was found to be as low as 0.02, and as this parameter increases, the oscillation mode stabilizes to a very specific value. 
Some numerical simulations of BNS mergers (see, e.g.~\cite{Soultanis_2022,DePietri_2020}) include evidence for the re-excitation of the $l=m=2$ $f$-mode in the remnant.

In this study, we focus on two different cases of post-merger waveforms from the study of
\cite{Soultanis_2022}, where the MPA1 EoS~\cite{mpa1_eos} was used. We explore whether rotational instabilities could be detected with future upgraded or 3rd-generation networks of detectors. For parts of our analysis, we use the \texttt{\textsc{BayesWave}} pipeline~\citep{Cornish_2015, Tyson_2015, Cornish_2020}, a Bayesian data-analysis algorithm useful to recover a GW signal via sine-Gaussian wavelets, which has already been extensively used for BNS post-merger studies (see, e.g., Ref.~\citep{Becsy_2017, Chatziioannou_2017, Torres_Rivas_2019,Chatziioannou_2021,Wijngaarden_2022, Miravet_2022}). We inject the simulated signals into colored Gaussian noise corresponding to different detectors. Our sources are considered to be optimally oriented with respect to the detector located in Livingston or with a non-optimal inclination and sky position at distances of at least\footnote{The lower limit of 40 Mpc is set due to the discovery of the GW170817 event.} 40 Mpc. In order to infer the capabilities of detecting rotational instabilities in BNS merger remnants, we calculate the overlap for the instability-part only. 

In the next sections, we demonstrate the effects of the low$-|T/W|$ rotational instability (Sec.~\ref{sec:waveform}) and define the different network configurations (Sec.~\ref{sec:network}), with which we aim to detect the instability. In Sec.~\ref{sec:injections}, the injections of our investigation are presented, and in Sec.~\ref{sec:method} a description of our analysis method is given, which is based on the \texttt{\textsc{BayesWave}} algorithm. The simulation setup and evaluation process are presented in Sec.~\ref{sec:evaluation}. Finally, our results are provided in Sec.~\ref{sec:results} and discussed in Sect.~\ref{sec:conclusions}.

\section{Post-merger waveform}
\label{sec:waveform}
In order to constrain the detectability of the low-$|T/W|$ rotational instability, we employ numerically generated post-merger waveforms, which reveal the re-excitation of the $m=2$ $f-$mode. More specifically, we study cases of NS with equal masses, which were simulated for the MPA1 EoS~\cite{mpa1_eos} using the \texttt{\textsc{Einstein Toolkit}} software~\cite{EinsteinToolkit:2022_11} in~\cite{Soultanis_2022}.  We focus on the cases of 1.50+1.50 $M_\odot$ and 1.55+1.55 $M_\odot$  equal-mass mergers.

\begin{figure}[htp]
\includegraphics[width=1\columnwidth]{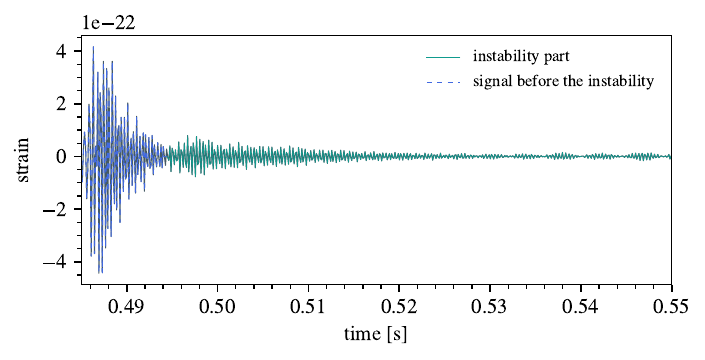}%

\includegraphics[width=1\columnwidth]{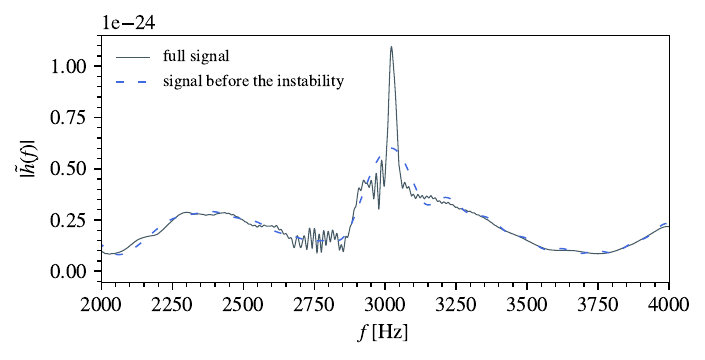}%
 	\caption{{\it Top panel}: GW strain of a BNS merger at 40 Mpc with the MPA1 EoS and a total mass of 3.1$M_\odot$ from (data from ~\cite{Soultanis_2022}). The instability part is shown in green. {\it Bottom panel}: The corresponding frequency-domain representation of the entire post-merger phase (solid line) and the signal before the instability (dashed line).} %The rotational instability appears as a narrow line on top of the spectrum generated in the early part of the post-merger phase.}
\label{fig:instability_representation}
\end{figure}

 Fig.~\ref{fig:instability_representation} shows the GW signal\footnote{The signal is projected onto the HF detector (see Sec. \ref{sec:network}) from an optimal orientation. The origin of time in Fig.~\ref{fig:instability_representation} is set to approximately half a second before the merger. } (gray) of a BNS merger at 40 Mpc with MPA1 EoS and a total mass ($M_{\rm{tot}}$) of 3.1 $M_\odot$ from~\cite{Soultanis_2022}. 
For the first $O(10)$ ms after merger, the GW signal is decaying. However, after this initial period, the rotational instability sets in
and causes a re-excitation, which maintains appreciable amplitude, showing only a slower damping rate with time. In the upper panel of Fig.~\ref{fig:instability_representation}, the waveform is split into the signal from the merger up to the onset of the instability (dashed blue curve) and the subsequent instability part (green), when the instability develops.  In the lower panel of Fig.~\ref{fig:instability_representation}, we observe that the Amplitude Spectral Density (ASD) of the full signal has an additional narrow peak, reaching about twice the ASD of the signal truncated at the time of the onset of the instability (dashed blue curve). The main focus of the present work is to investigate whether the presence of such an additional narrow peak could become detectable with different GW detector networks. 

\section{Network Configuration\label{sec:network}}

Our study focuses on the detectability of the low-$|T/W|$ rotational instability in post-merger NSs by future observatories. For that reason, we consider three cases of network configuration, for which a summarized description is given in Table \ref{tab:network_conf}. The first network configuration corresponds to {\it twice} the projected sensitivity of the fifth planned observing run (O5) Advanced Virgo~\citep{Virgo, O5_curves}, and $A+$ (LIGO)~\citep{LIGO, O5_curves}). 

\begin{table}[htp]
  \centering
    \begin{tabular}{l@{\hskip 0.4in}c@{\hskip 0.2in}c}
    \hline
    \hline
      \textbf{Label} & \textbf{Detectors} & \textbf{Sensitivity} \\
      \hline
    \multirow{2}{*}{$2~\times~\rm{O5}$} & 2 LIGOs (H, L) & $2~\times~\rm{A+}$ \\ % <-- Combining 2 rows 
    & Virgo (V) & 2 x AdV (O5)\\
    \hline
      \multirow{2}{*}{CE+ET} & Cosmic Explorer (CE) & CE-20-pm \\ % <-- Combining 2 rows 
      & Einstein Telescope (ET) & ET-D \\
      \hline
      \multirow{2}{*}{HF} & 25 km L-resonator & \multirow{2}{*}{HF}\\
      & Interferometer\\
      \hline
      \hline
    \end{tabular}
    \caption{Different network configuration or detector designs considered in this work. }
    \label{tab:network_conf}
\end{table}

The second network is more sensitive for post-merger NS signals and consists of the planned Cosmic Explorer (CE)~\citep{CE_curves, CE_science} and the planned Einstein Telescope (ET)~\citep{ET_curves}, comprising the 3G detectors, to become operational within the next decade. For both detectors, there have been different estimates for their sensitivities, but in this study we choose to work with the ``CE 20 km" for the post-merger curve~\citep{CE_curves, CE_curves_site} and the ``ET-D"~\cite{ET,ET_curves}, because these designs have greater sensitivity to post-merger NS signals. Regarding their locations, the CE detector is assumed to be located in Livingston and ET in Cascina~\cite{CE_ET_locations}. 

The third configuration refers to the recent proposal for a 25 km L-resonator interferometer~\cite{HF}, denoted as HF (High Frequency), which we assume the LIGO-Livingston location. A significant limitation in future detectors, which are based on dual-recycled-Fabry-Pérot-Michelson Interferometers, is the loss in the signal extraction cavity (SEC). The HF design aims to suppress this loss-limit at high frequencies, resulting in a larger than 1/yr detection rate of the post-merger signal~\cite{HF}, significantly better than other 3G detector designs. The HF design includes an ``L" shape optical resonator of 25 km arm length and laser wavelength at 1064 nm (for a detailed description, see Sec. IV in Ref.~\cite{HF}). Their proposed configuration targets post-merger signals of BNSs (sensitive between 2-4kHz) with a peak sensitivity at 3kHz. If realized, the HF detector would be an ideal interferometer for investigations focusing on the NS post-merger signal, since the peak frequency of such signals is between 2kHz and 4kHz. For our chosen models, the peak frequency of the post-merger signal is close to 3 kHz~\cite{Soultanis_2022}. 

We investigate source cases that are either optimally oriented with respect to the Livingston detector or randomly oriented. 
The \textit{antenna pattern response functions} $F_+, \,F_\times$
of the interferometer as defined in the sky plane, are
\begin{equation}
\begin{aligned}
F_{+,\times} =& -\frac{1}{2}\big(1+\cos^2\theta\big)\cos2\phi \cos2\psi \\ 
& \mp \cos\theta\sin2\phi \sin2\psi,  
\end{aligned}
\label{eq:responses}
\end{equation} 
with ($\theta, \, \phi$) denoting the sky-location of a source, relative to the axes of the detector arms. The two polarization components $h_+$ and $h_\times$ are with respect to axes in the plane of the sky that are rotated by an angle $\psi$ (see Fig. 3 of Ref.~\cite{Sathyaprakash_2009}). Then, one can define the right ascension of the source, $\alpha=\phi +$GMST, and its declination, $\delta=\pi/2 - \theta$, where GMST is the Greenwich mean sidereal time of arrival of the
signal. Therefore, the quantities {$\alpha,~\delta$} maximize the responses {$F_+,F_{\times}$}, when $\phi$ is equal to the detector longitude ($-1.58430937078$ rad at the location of the LIGO Livingston detector) and $\theta$ the detector latitude ($0.53342313506$ rad). Furthermore, we assume that \textit{the binary is seen face-on}, i.e. the inclination is $\iota=0$ (see \cite{geometry}). With these assumptions, the source is considered to be optimally oriented with respect to the detector. 

\section{Injections}
\label{sec:injections}

For each BNS merger case, we construct two different source cases to perform the injection, differing only in their sky localization and orientation. One case has a non-optimal orientation and inclination (using the inferred values for the GW170817 event \cite{GW170817, GW170817_multi, GW170817_sky_param, i_GW17}). The second case is an optimally oriented source with respect to the LIGO-Livingston detector (denoted as L). The parameters of the injected model are given in Table \ref{tab:injected_models}. The distance from the source depends on the detectability of the network configuration (see Table~\ref{tab:network_conf}) and ranges from 40 $\mrm{Mpc}$ to 200 $\mrm{Mpc}$. The lower limit of 40 $\mrm{Mpc}$ is set due to the inferred distance of GW170817.

We assume two different cases of total mass $M_{\rm{tot}}$ 3.0 and 3.1 $M_\odot$. In Ref.~\cite{Soultanis_2022}, a re-excitation of the GW signal is demonstrated, which is a characteristic of the presence of rotational instabilities. 
Although the inclusion of magnetic fields would affect the rotational profile, as a first step, we consider simulations without magnetic fields or other viscous effects.

\begin{table}[H]
  \centering
    \begin{tabular}{l@{\hskip 0.2in}c@{\hskip 0.18in}c}
    \hline
    \hline
       \multirow{2}{*}{\textbf{Extrinsic Parameters}} & \textbf{Case 1:} & \textbf{Case 2:} \\ % <-- Combining 2 rows 
       &  \textbf{Non-optimal} & \textbf{Optimal}\\
      \hline
    GPS trigger time ($t_c$) & 1187008882.43  & 1187008882.43 \\
    Inclination & 0.5585 & 0 \\ 
    Right Ascension [rad] & 3.44616 & 1.14469\\
    Declination [rad] & -0.408084 & 0.53342 \\
    \hline
    \hline
    \end{tabular}
    \caption{Extrinsic parameters for two difference cases: non-optimal orientation and optimal orientation with respect to the LIGO Livingston detector.}
    \label{tab:injected_models}
\end{table}

The duration of the simulated signal is 74~$\mathrm{ms}$, which is then injected into 1 second of colored Gaussian noise. The frequency band of the analysis is (1800, 4096)\footnote{The value of 4096Hz is the Nyquist frequency, for our sampling rate of 8192Hz.} Hz for the case of $M_{\rm{tot}}=3.0M_\odot$ and (2000, 4096) Hz when $M_{\rm{tot}}=3.1M_\odot$. The lower cutoff frequency is chosen with the aim of leaving out parts of the spectrum that are influenced by the inspiral signal.

\section{Data Analysis methodology\label{sec:method}}

In order to assess the detectability of the various signals that appear with different characteristic spectra, we adopt a template-agnostic method in a Bayesian framework. In this framework one begins with the \textit{Bayes’ Theorem}, which is expressed as
\begin{equation}
p(\param|s, \mathcal{M})=\frac{p(s|\param, \mathcal{M})p(\param, \mathcal{M})}{p(s, \mathcal{M})},
    \label{Bayes theorem}
\end{equation}
where $p(\param|s, \mathcal{M})$ is the posterior for the parameters $\param$ of the signal $h$ given the data $s$ and model $\mathcal{M}$; $p(\param, \mathcal{M})$ is the adopted prior of the parameters of interest; and  $p(s|\param, \mathcal{M})$ is the likelihood function of the data, given the particular signal. $p(s, \mathcal{M})$ is the marginalized likelihood, or \textit{evidence} (integral of the likelihood function over the parameter space) 
\begin{equation}
\begin{aligned}
    \mathcal{Z}&=\int p(\param,s|\mathcal{M})d\param\\
&=\int p(s|\param, \mathcal{M})p(\param| \mathcal{M})d\param,
\end{aligned}
    \label{evidence}
\end{equation}
which, in parameter estimation procedures serves solely as a normalizing constant~\cite{gelmanbda04}.

However, the evidence is crucial for model selection purposes, because it essentially characterizes the capabilities of our adopted model to describe the measurements~\cite{silvia, Cornish_2015, Chatziioannou_2017}. Having the evidences of two competing models, one can compute the Bayes Factor as the ratio of the evidences and infer which model is supported best by the measured data. In practice, the integral of Eq.~(\ref{evidence}) is quite challenging to compute directly, but one can adopt numerical methodologies to approximate it (e.g. via Thermodynamic Integration~\cite{thermodynamic_int} or for different approaches, see~\cite{Stepping-stone}). However, there are situations where the set of models to be tested is relatively large, and computing the evidence for each combination of models can be quite inefficient or even computationally prohibiting. Instead, a trans-dimensional Markov Chain Monte Carlo (MCMC) method can be adopted, which would yield an estimate of the Bayes factor directly~\cite{Green,Cornish_2015,Eryn}).

\begin{figure}[htp]
  \subfloat[A low network SNR of 11.]{%
\includegraphics[clip,width=0.75\columnwidth]{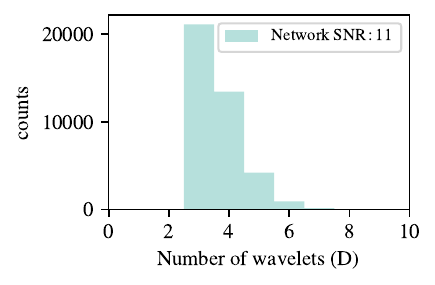}%
}

    \subfloat[A high SNR of 134.]{%
\includegraphics[clip,width=0.75\columnwidth]{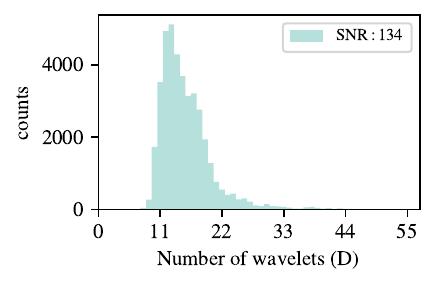}%
}
 	\caption{Histogram of the number of wavelets D used by \texttt{\textsc{BayesWave}} to reconstruct the injected entire post-merger GW signal for the case of an optimally-oriented BNS merger at 40 Mpc, with the MPA1 EoS and $M_{\rm{tot}}=3.1$ \(M_\odot\). {\it Top panel}: For the network configuration with a sensitivity twice the expected sensitivity of the O5 run, the SNR of the entire post-merger signal is 11 and it is reconstructed with a small number of wavelets (but with larger uncertainties). {\it Bottom panel}: For the HF detector design, the SNR of the entire post-merger signal is 134 and a larger number of wavelets leads to smaller uncertainties in the reconstruction.}
 \label{fig:D}
\end{figure}

For the reason mentioned above, in this work we adopt the \texttt{\textsc{BayesWave}} pipeline~\citep{Cornish_2015, Tyson_2015}, which utilizes Bayesian inference to sample a dynamical parameter space. In practice, signal and noise are modeled by an ensemble of Morlet–Gabor wavelet functions by employing a Reversible-Jump (RJ) MCMC~\cite{RJMCMC} algorithm, where the \textit{optimal} number of wavelets and their corresponding parameters are estimated from the data. 
In fact, the more complex the data, the larger the number of wavelets is, as in Fig.~\ref{fig:D}. By ``optimal'', we refer to the statistically most probable model that can sufficiently describe the observations. Essentially, Occam's razor is applied on the model complexity, given the data~\cite{Chatziioannou_2017, Yi_Millhouse}. The same approach is used to separate the noise (together with possible non-stationarities, such as glitches~\cite{Hourihane_2022}) and the gravitational-wave signal components. 

\begin{figure}[htp]
\includegraphics[width=1\columnwidth]{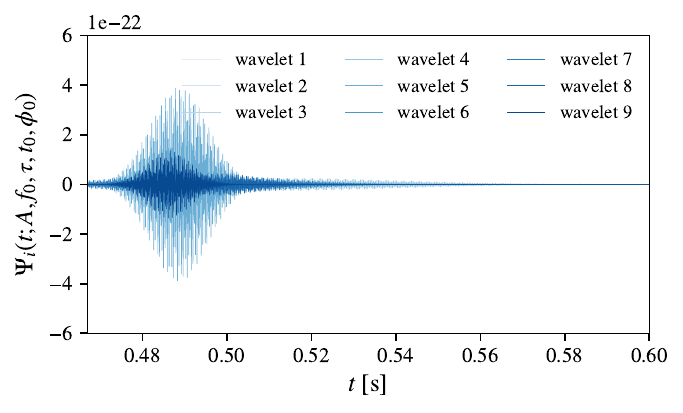}
\caption{The set of wavelets $\Psi_i$  used by \texttt{\textsc{BayesWave}} to reconstruct the entire post-merger signal $h_{+}(t)$ of an optimally-oriented BNS merger at 40 Mpc, with the MPA1 EoS and a total mass of 3.1 \(M_\odot\) for HF detector design.  A specific sample from the MCMC chains was chosen, which yields a signal reconstruction using 9 wavelet components.}
 \label{fig:psi_i}
\end{figure}

As already mentioned, \texttt{\textsc{BayesWave}} uses a more flexible parameterization for the signal $h$. The signal is modeled as a sum of functions $\Psi_i(\param)$, each depending on a set of parameters $\param$. Then, $\Psi_i$ can be written as~\cite{Cornish_2015},
\begin{equation}
\begin{aligned}
\Psi(t;A, f_0, \tau, t_0, \phi_0)=&Ae^{-(t-t_0)^2/\tau^2}\\
&\times\cos[2\pi f_0(t-t_0)+\phi_0],
\end{aligned}
    \label{eq:wavelet basis}
\end{equation}
where $t_0,\, f_0$ are the central time and frequency of the wavelet, respectively. In Eq.~(\ref{eq:wavelet basis}), $A$ is the wavelet amplitude, $\phi_0$ a phase offset and 
\begin{equation}
    \tau = \frac{Q}{2\pi f_0},
    \label{eq:tau}
\end{equation}
with Q being the quality factor\footnote{The quantity Q gives a sense of how localize in time the wavelet is.}. Thus, the GW signal at the geocenter is represented as 
\begin{equation}
h_+(t) = \sum_{i=0}^D\Psi_i(t;A, f_0, \tau, t_0, \phi_0),
    \label{eq:hp}
\end{equation}
where $D$ is the number of wavelets. An example of such a representation is given in Fig.~\ref{fig:psi_i}, where the ensemble of wavelets $\Psi_i$ is to be used in order to obtain the $h_{+}(t)$ reconstruction.
Continuing in the frequency domain, we write  $\tilde{h}_\times(f)$ as
\begin{equation}
\tilde{h}_\times(f) = \epsilon \tilde{h}_+(f) e^{i \pi/2},
    \label{eq:hc}
\end{equation}
where $\epsilon$ is the ellipticity\footnote{See a detailed review analysis for ellipticity and polarizations in~\cite{Maximiliano_Isi_review}.} parameter~\cite{Cornish_2015}. For a linearly-polarized
wave, $\epsilon=0$, while $\epsilon=1, -1$ yields circular polarization. The expression in Eq.~(\ref{eq:hc}) is a consequence of modeling GW signals as a superposition of the elliptical state, which allows us to encapsulate all possible morphologies of a fully polarized monochromatic wave. By decomposing the signal in terms of spherical harmonics $_{-2}Y_{lm}$, one can obtain the ellipticity parameter relative to the inclination $\iota$ of the source as in~\cite{Maximiliano_Isi_review}
\begin{equation}
\epsilon =  \frac{2\cos\iota}{1+\cos\iota^2}.
    \label{eq:epsilon relative to iota}
\end{equation}

Then, \textit{the strain in the frequency domain measured by the detector} is expressed in the form~\cite{Thorne}
\begin{equation}
\tilde{h}(f) = \Bigg[\tilde{h}_+(f)F_+ +\tilde{h}_\times(f)F_\times\Bigg] e^{i 2\pi\Delta t(\alpha, \delta)+i\phi},
    \label{eq:hf reconstruction}
\end{equation}
where $\Delta t(\alpha, \delta)$ is the time-delay relative to the arrival time at the geocenter and $\phi$ the phase at a fiducial reference time. 

To calculate the reconstructed signal using  Eq.~(\ref{eq:hf reconstruction}), we need to estimate both extrinsic and intrinsic parameters. These are the sky position, orientation, and number of wavelets $(\alpha, \, \sin\delta, \, \psi, \, \cos\iota, \, \phi, \, D)$, and the parameters of each wavelet $(A, \, f_0, \, \tau, \, t_0, \, \phi_0)$, respectively. As mentioned above, these are dynamically sampled with the RJMCMC algorithm. We employ uninformative priors,  assuming that $t_{0}\in \mathcal{U}[t_{c}-0.5,t_{c}+0.5]$, where $t_{c}$ is the GW trigger time, and $\phi_{0}\in \mathcal{U}[0,2\pi]$. For the amplitude, we choose the default prior suggested in~\cite{Cornish_2015}. Regarding the quality factor $Q$, its maximum value is set to $300$ for all network configurations, except the HF detector. For this detector, we set it to 800 due to its improved sensitivity, which results in a relatively high post-merger SNR and consequently in a higher quality factor. Finally, we employ a prior for the dimensionality analogous to \cite{Chatziioannou_2017}, i.e. $D_{\rm{min}}=2$.
Furthermore, we set the number of iterations to $4\times10^6$, of which half are considered a burn-in period and later discarded. The final chain is additionally thinned, by keeping every 100th sample. Thus, we calculate $N$=20000 reconstructions. 

As a measure to quantify the quality of the reconstructions, we use the \textit{overlap} with respect to the injected signal. The overlap between the \textit{true injected signal} in the frequency domain $\tilde{h}_s(f)$ and the \textit{reconstructed} $\tilde{h}_r(f)$ is obtained via 
\begin{equation}
\mathcal{O}=\frac{\langle \tilde{h}_s,\tilde{h}_r \rangle}{\sqrt{\langle \tilde{h}_s,\tilde{h}_s \rangle}\sqrt{\langle \tilde{h}_r,\tilde{h}_r \rangle}},
\label{eq:overlap}
\end{equation}
where the $\bm{\langle \cdot \, , \cdot \rangle}$ represents the inner product between two real time series. This is defined as~\cite{Maggiore:2007ulw}
\begin{equation}
\langle a,b \rangle=2 \int_{f_\mrm{low}}^{f_\mrm{high}}\frac{\tilde{a}(f)\tilde{b}^*(f)+\tilde{a}^*(f)\tilde{b}(f)}{S_n(f)} d f , 
\label{eq:inner_product}
\end{equation}
with $S_n(f)$ being the detector's one-sided noise Power Spectral Density (PSD) and $(f_\mrm{low} , f_\mrm{high})$ the bandwidth of the analysis, while the asterisk $(^\ast)$ denotes the complex conjugation. The overlap takes values between -1 and 1 ~\citep{Cornish_2015, Maximiliano_Isi_review}; the closer the quantity to 1, the better the reconstruction, while $\mathcal{O}$ = -1 means an anti-correlation. Note that the optimal signal-to-noise ratio (SNR) is defined as
\begin{equation}
\mrm{SNR}=\sqrt{\langle \tilde{h}_s,\tilde{h}_s \rangle}.
\label{eq:SNR equation}
\end{equation}

\begin{figure}[htp]
    \subfloat[Frequency-domain representation.]{%
\includegraphics[clip,width=0.95\columnwidth]{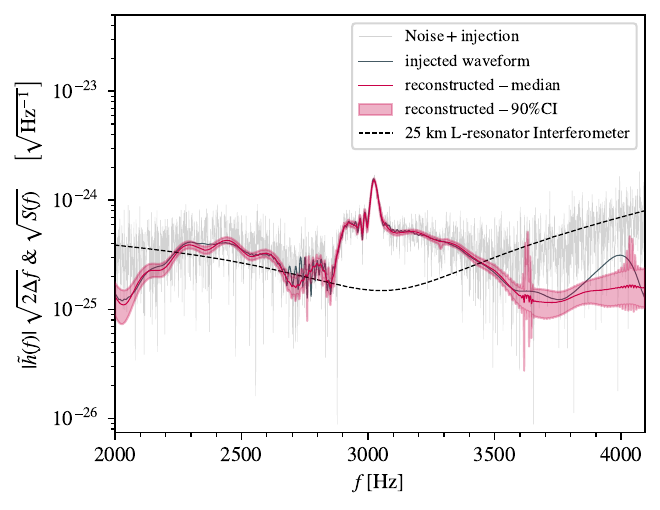}%
}

    \subfloat[Time-domain representation.]{%
\includegraphics[clip,width=0.95\columnwidth]{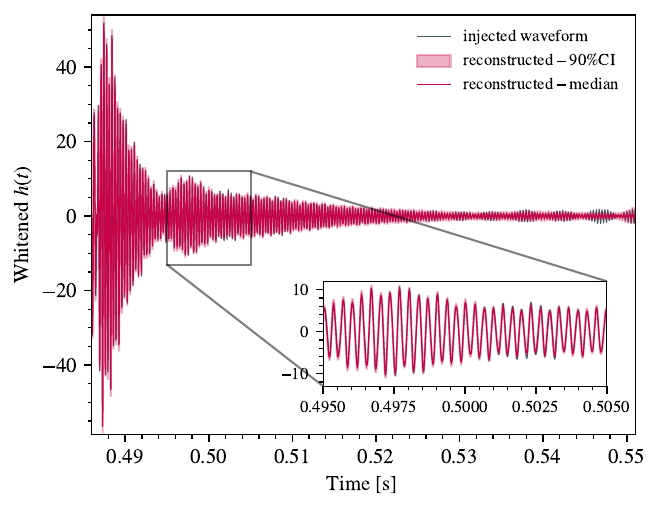}%
}
 	\caption{Reconstruction of the entire post-merger phase for an optimally-oriented BNS merger with EoS MPA1 and $M_{\rm{tot}}=3.1~M_\odot$, for the HF detector design. {\it Top panel}: Recostruction in the frequency domain. {\it Bottom panel}: Recostruction in the time domain.}
 \label{fig:TD_and_FD_full_signal_40_MPA1_155155}
\end{figure}

\section{Evaluation method}
\label{sec:evaluation} 
The purpose of this work is to assess the detectability of the rotational instabilities in the post-merger GW spectrum. We anticipate that all networks (Table~\ref{tab:network_conf}) will accurately detect and reconstruct the signal that corresponds to the ``merger" and ``early" post-merger phases (up to $O(10)$ms after merger, where the main post-merger GW emission takes place), but not to the ``late" post-merger  signal (i.e. to the emission after $O(10)$ms from merger), where rotational instabilities could trigger a re-excitation of GW emission. It is essential to note that the overlap of the entire post-merger signal will be rather close to one in all cases considered here, since the numerator in Eq.~(\ref{eq:overlap}) is  dominated by the strong merger and early post-merger part of the waveform. To avoid biased results (in terms of the ability to infer the presence of rotational instabilities), we specifically constrain the overlap calculation to only the \textit{instability part} of the signal.

First, we used the \textsc{\texttt{BayesWave}} pipeline to analyze the entire post-merger signal, and obtain its extrinsic and intrisic parameters (described in Sec.~\ref{sec:method}). As an example, Fig.~\ref{fig:TD_and_FD_full_signal_40_MPA1_155155}a shows the resulting reconstruction of the entire post-merger signal in the frequency domain, for the case of a BNS source with $M_{\rm{tot}}=3.1~M_\odot$ at 40 Mpc, optimally oriented with respect to the HF detector. The corresponding reconstruction in the time domain is shown in Fig.~\ref{fig:TD_and_FD_full_signal_40_MPA1_155155}b. Since this is the most sensitive detector we consider and the source is placed at the nearest realistic distance, the SNR of the post-merger phase was 133, resulting in a very accurate reconstruction up to several tens of milliseconds after merger, including the instability part. In the inset of Fig.~\ref{fig:TD_and_FD_full_signal_40_MPA1_155155}b, we show in more detail a magnification of the beginning of the instability part, where the accuracy of the reconstruction can be clearly seen.

\begin{figure}[t]
\includegraphics[clip,width=1\columnwidth]{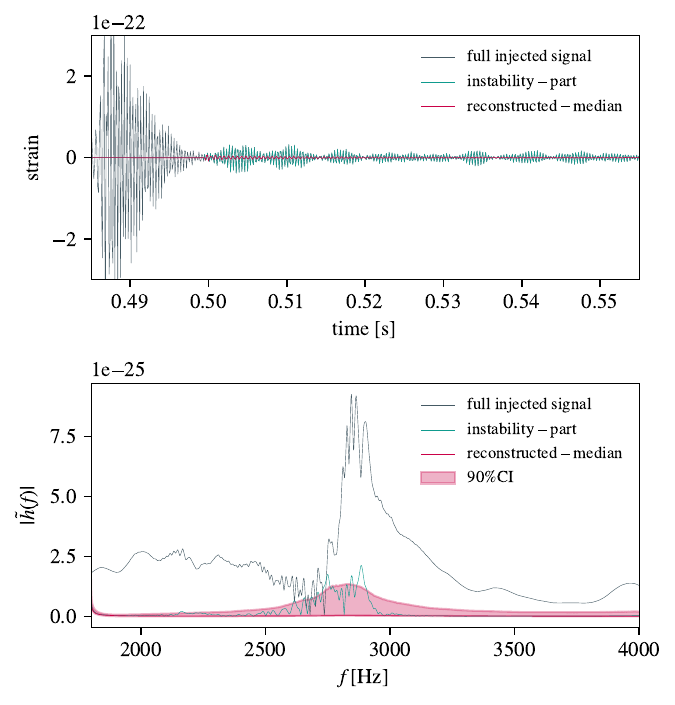}%
 	\caption{Reconstruction of the instability part of the post-merger signal for an optimally-oriented source at 40 Mpc with EoS MPA1 and and ${M_{\rm{{tot}}}=3.0  M_\odot }$ for the CE+ET detector network. {\it Top panel}: GW strain in the time domain, showing the different parts of the post-merger signal and the reconstruction of the instability part. {\it Bottom panel}: The corresponding frequency-domain representation.} 
 \label{fig:cropped_CEET_150}
\end{figure}

\begin{figure}[t]
\includegraphics[clip,width=1\columnwidth]{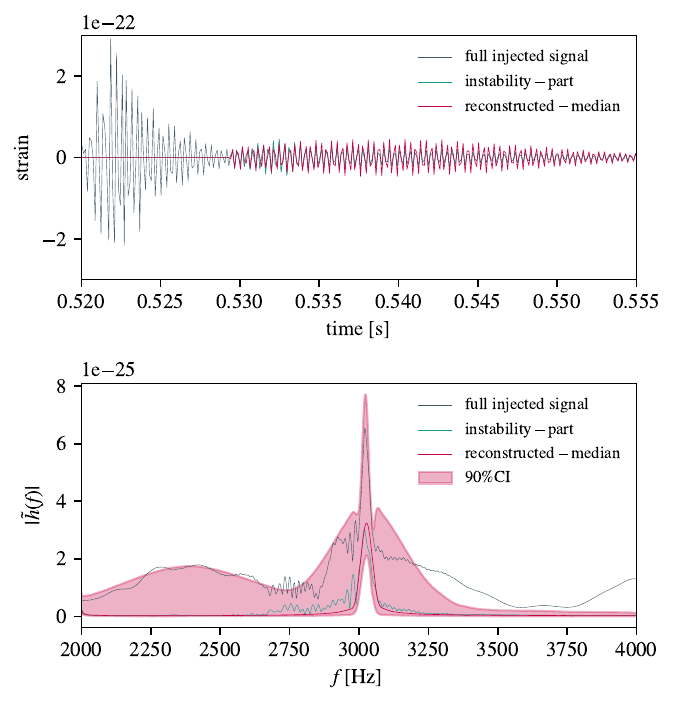}%
 	\caption{Same as Fig. \ref{fig:cropped_CEET_150}, but for a non-optimally oriented source with ${M_{\rm{{tot}}}=3.1 } M_\odot$.}
 \label{fig:cropped_CEET_155}
\end{figure}

\begin{figure}[t!]
  \subfloat[Source optimally-oriented at 40 Mpc.]{%
\includegraphics[clip,width=1\columnwidth]{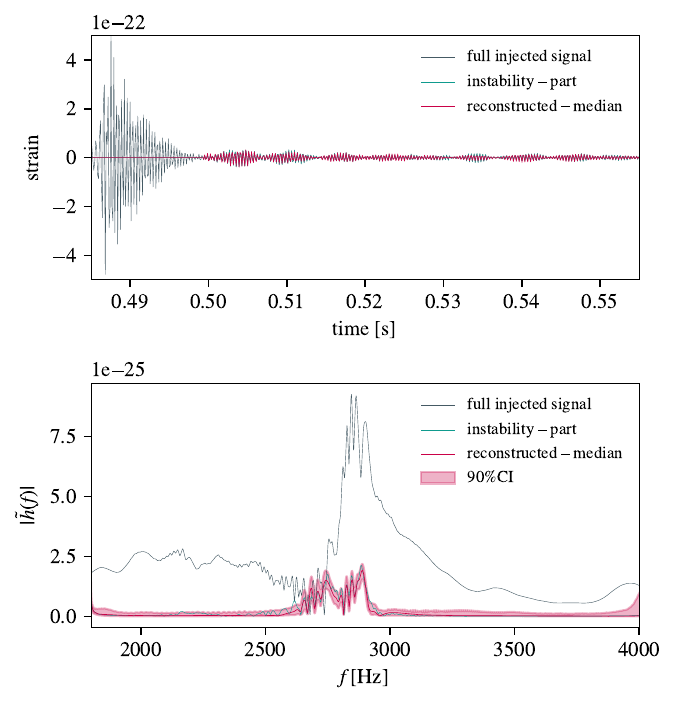}%
}

    \subfloat[Source optimally-oriented at 80 Mpc.]{%
\includegraphics[clip,width=1\columnwidth]{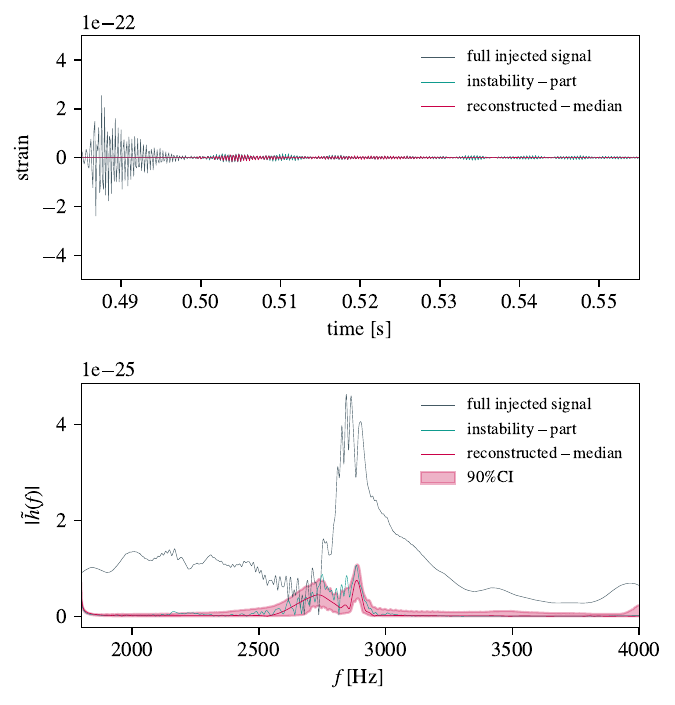}%
}
 	\caption{Same as Fig. \ref{fig:cropped_CEET_150}, but for the HF detector design and an optimally-oriented source at a distance of 40Mpc ({\it top panel}) and 80Mpc ({\it bottom panel}).}
 \label{fig:cropped_HF_150}
\end{figure}

\begin{figure}[t!]
  \subfloat[Source optimally oriented at 40 Mpc.]{%
\includegraphics[clip,width=1\columnwidth]{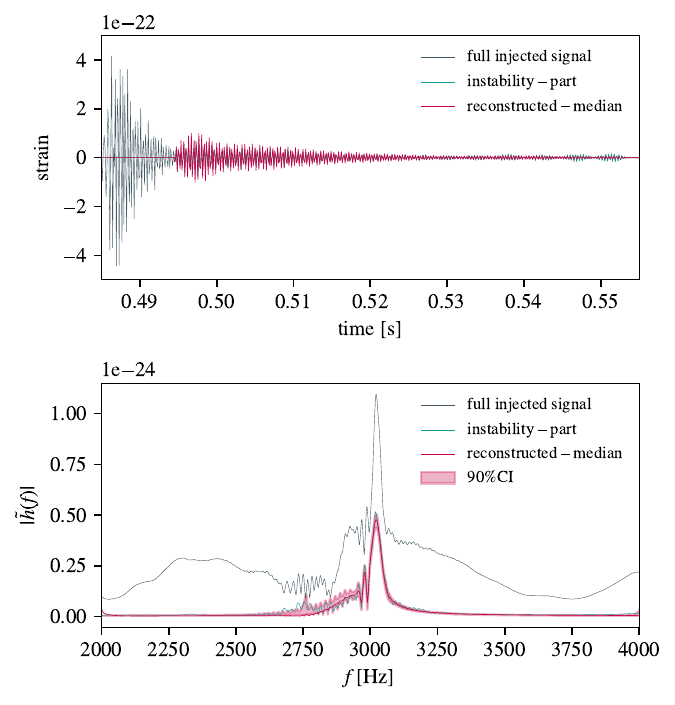}%
}

    \subfloat[Source optimally oriented at 200 Mpc.]{%
\includegraphics[clip,width=1\columnwidth]{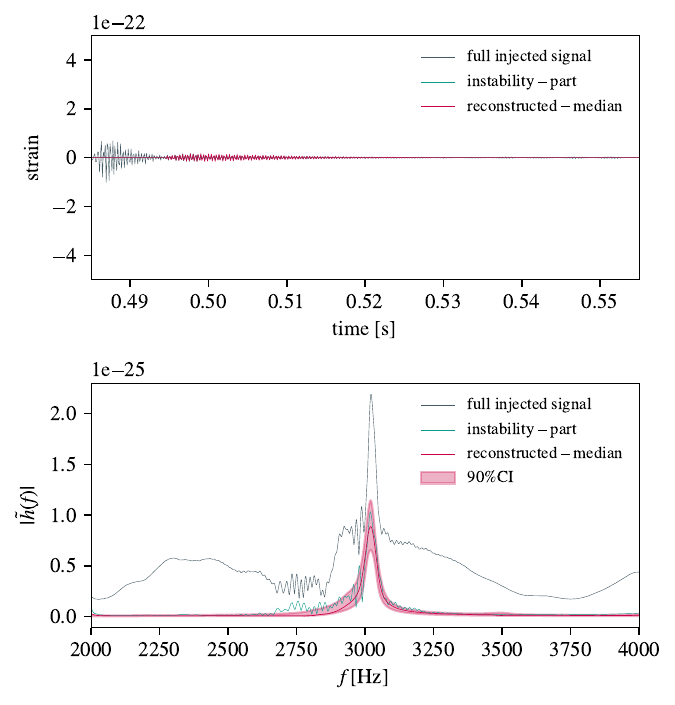}%
}
 	\caption{Same as Fig. \ref{fig:cropped_CEET_150}, but for a total mass of ${M_{\rm{{tot}}}=3.1} M_\odot$ and a distance of 40Mpc ({\it top panel}) and 200Mpc ({\it bottom panel}).}
 \label{fig:cropped_HF_155}
\end{figure}

\section{Detectability of the rotational instabilities\label{sec:results}}

Next, we isolated and evaluated the detectability of the instability part of the signal. To do this, we cropped the time series (around 0.4950s as shown in Fig.~\ref{fig:TD_and_FD_full_signal_40_MPA1_155155}b) and retained only the part of the signal at $t>0.4950s$, which contains the instability part. We applied a very weak Tukey window ($a_{\rm{wt}}=0.01$) and resized the signal onto the initial duration
(1 second) and re-sampled them to the desired sampling rate (8192 Hz). The same was applied to the 20000 samples of the posterior reconstructed signal and a new median of the reconstructed instability part was computed and used in evaluating the overlap restricted to the instability part. 
The detailed results of this investigation are presented in Sect.~\ref{sec:results}. In Appendix~\ref{sec:appendixb}, we present additional cases (for the same model, but different detector networks), where we analyzed the entire post-merger signal. 

\subsection{Reconstruction of the instability part\label{sec:recovered}}

Following the procedure described in Sec. \ref{sec:evaluation}, we reconstructed the instability part of the post-merger signals for the different cases we considered. 

The first network configuration in Table~\ref{tab:network_conf}, corresponding to possible future upgrades of the HLV network at twice the sensitivity of O5, did not have the required sensitivity for the instability part to be reconstructed, for any of the injections we considered. At this sensitiviy, only the initial post-merger signal can be reconstructed (see Appendix~\ref{sec:appendixb}).

Next, we consider the CE+ET network configuration in Table~\ref{tab:network_conf}, for the case of 
an optimally oriented source at a distance of 40Mpc, with $M_{tot}=3.0 M_\odot$. The top panel of Fig.~\ref{fig:cropped_CEET_150} shows, in the time domain, the injected signal for the entire post-merger phase, the instability part and its corresponding median reconstructed signal. The corresponding frequency-domain representation is shown in the bottom panel of the same figure. For this case, the instability part has a small amplitude and the median reconstructed signal has a very small overlap with the injected data. The case of non-optimal orientation is even less promising. Hence, for this configuration, one would not be able to discern the presence of a rotational instability. 

For the same CE+ET network configuration, we also consider the case of a merger with $M_{\rm{tot}}=3.1 M_\odot$. In this case, we do not assume optimal orientation and the distance is  40 Mpc. The top panel of Fig.~\ref{fig:cropped_CEET_155} shows that the instability part is reconstructed in the time domain, although the median waveform has a larger amplitude than the injected data. In the bottom panel of Fig.~\ref{fig:cropped_CEET_155}, which displays the corresponding picture in the frequency domain, one can observe that the 90\% confidence interval (CI) region is rather broad, but the instability part is still well within this region and with a somewhat lower amplitude than the median reconstructed Fourier transform $\tilde h(f)$.  We conclude that, for this case, one would be able to infer the presence of the rotational instability, but the parameter estimation would not be very accurate.

Finally, we consider the HF detector design in Table~\ref{tab:network_conf}. The two panels of Fig.~\ref{fig:cropped_HF_150}a show that for an optimally-oriented source with $M_{\rm{tot}}=3.0 M_\odot$ at a distance of 40Mpc, the instability part can be reconstructed fairly accurately in the time and frequency domains. Increasing the distance to 80Mpc (two panels of Fig.~\ref{fig:cropped_HF_150}b) results in a similar median reconstructed signal, but with a broader 90\% CI that would not allow an accurate parameter estimation. For both distances of 40Mpc and 80Mpc, the reconstruction reveals the presence of two frequencies with a separation of about 150Hz, which explains the apparent ``beating'' in the time domain representation. One frequency has a value comparable to the dominant $f_{\rm peak}$ frequency in the early post-merger phase, whereas the other frequency is somewhat smaller. The two distinct frequencies have similar amplitudes in the frequency domain, indicating that they both persist throughout the duration of the signal in the instability part. Hence, they could be due to two different unstable modes operating at the same time. This remains to be confirmed with a detailed mode analysis.

For the case of $M_{\rm{tot}}=3.1 M_\odot$ the HF detector achieves a very accurate reconstruction, when the source is optimally-oriented at 40Mpc, as shown in the two panels of Fig.~\ref{fig:cropped_HF_155}a. For this case, only a single frequency (comparable to the dominant $f_{\rm peak}$ frequency in the early post-merger phase) is active in the instability part and it is recovered with a very narrow 90\% CI in the frequency domain, which would allow for accurate parameter estimation. When the same source is set at a significantly larger distance of 200Mpc, the reconstruction is still fairly accurate, but with somewhat broader 90\% CI.

\subsection{Overlap calculations for the instability part\label{sec:performance}}

\begin{figure}[htp]
  \subfloat[$M_{\rm{tot}}=3.0$ $M_\odot$.]{%
\includegraphics[clip,width=0.67\columnwidth]{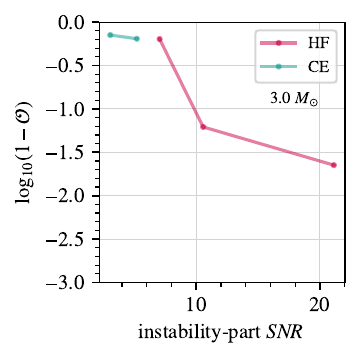}%
}

    \subfloat[$M_{\rm{tot}}=3.1$ $M_\odot$.]{%
\includegraphics[clip,width=0.67\columnwidth]{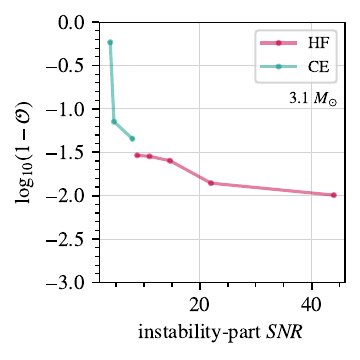}%
}
 	\caption{Overlap of the median reconstructed instability part of the post-merger signal for a BNS merger with
  {(a)} $M_{\rm{tot}}=3.0$ $M_\odot$ and  {(b)} $M_{\rm{tot}}=3.1$ $M_\odot$. The overlap obtained for the {CE}  and {HF} detectors is shown.}
\label{fig:overlap_instability_part}
\end{figure}

Fig. \ref{fig:overlap_instability_part}a shows the calculated overlap between the median reconstructed signal at the CE detector and the injected signal for the instability part as a function of the SNR of this part. The model in this case is the $M_{\rm }=3.0M_\odot$ model and we varied the distance to the source to achieve different SNR values. For this model, the instability part is very weak and with the CE+ET network only a very small overlap of ${\cal O} \simeq 0.36$ is achieved. For the same model, the HF detector design performs significantly better, as one can reach ${\rm SNR}\sim 21$ for the instability part of the signal, with an overlap of ${\cal O} \simeq 0.977$.

In \ref{fig:overlap_instability_part}b the corresponding overlap for the case of $M_{\rm }=3.1M_\odot$ is shown. This model has a significantly stronger instability signal, allowing the CE+ET network to reach an ${\rm SNR}\sim 8$ with an overlap of ${\cal O} \simeq 0.96$. On the other hand, the HF detector reaches an ${\rm SNR}\sim 44$ with an overlap of ${\cal O} \simeq 0.99$.

\begin{figure}[htp]
  \subfloat[CE detector performance regarding the instability.]{%
\includegraphics[clip,width=0.67\columnwidth]{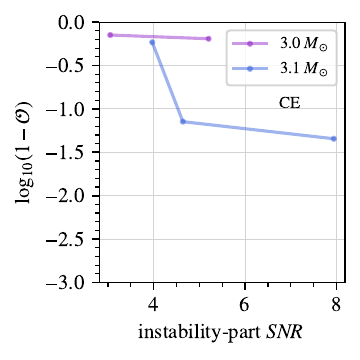}%
}

    \subfloat[HF detector performance regarding the instability.]{%
\includegraphics[clip,width=0.67\columnwidth]{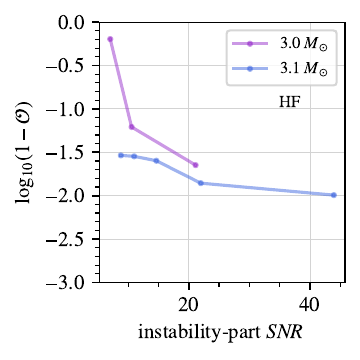}%
}
 	\caption{Same as Fig. \ref{fig:overlap_instability_part}, but the comparison is now between the two different mass cases for a) the CE dector and b) the HF detector.}
\label{fig:overlaps_instability_CE_ET}
\end{figure}

To allow a direct comparison between the $M_{\rm }=3.0M_\odot$ case and the $M_{\rm }=3.1M_\odot$ case, we display the overlap obtained at the CE detector and with the HF detector design in Figs. \ref{fig:overlaps_instability_CE_ET}, respectively.

\section{Summary and Conclusions\label{sec:conclusions}}
Several studies on the dynamical evolution of differentially rotating neutron star configurations suggest a prospective excitation of a low-$|T/W|$ rotational instability, which could play an important role in the long-term evolution of a BNS post-merger remnant \citep{Passamonti_2020, Centrella_2001, Watts_Andersson_Jones}. This phenomenon causes a dynamical shear instability \citep{Passamonti_2020, Watts_Andersson_Jones}, while stars with a large entropy could be stabilized against it \citep{Camelio_2021}. Evidence for a re-excitation of the of the $m=2$ $f$-mode can be seen in the numerical simulations of e.g. \cite{DePietri_2020,Soultanis_2022,Xie}. 

We explored the potential for detecting such rotational instabilities if they occur in BNS merger remnants. To do that, we inject the numerically simulated signals of equal-mass BNS mergers produced by \cite{Soultanis_2022}, with MPA1 EoS and total mass of 3.0 and 3.1 $M_\odot$, into different network configurations (see Table~\ref{tab:network_conf}) assuming colored Gaussian noise. We considered that the merge occurs at realistic distances, that is, 40-200 Mpc, and examined different source orientations (see Table~\ref{tab:injected_models}). Finally, we employed \textsc{\texttt{BayesWave}} to compute N=20000 reconstructed signals, and then we calculated the overlap between the injected waveform and the median reconstructed of the instability part of the signal.

Our study indicates that a network of  advanced detectors upgraded to twice the   sensitivity of the A + design would only be able to detect the ``early" post-merger signal at the closest distance of 40 Mpc, but it would miss the presence of rotational instabilities. 
We find that for a BNS with $\rm{M_{tot}}=3.1$ \(M_\odot\) using the MPA1 EoS, the network comprised of the proposed CE and ET detectors will be able to detect rotational instabilities if the event takes place at distances less than 80 Mpc. The HF design would detect rotational instabilities for sources within $\sim O$(200Mpc).

For the case of a BNS merger with MPA1 EoS and a lower mass of $\rm{M_{tot}}=3.0$ $M_\odot$, all considerd detector designs reconstruct the early post-merger signal of the BNS merger, but not necessarily the signal regarding rotational instabilities. In this case, the spectrum of the waveform reveals the presence of two frequencies, the dominant mode and a second, nearby frequency, which have comparable amplitudes in the frequency domain. 
For this model, the CE+ET network would not allow for an accurate parameter estimation of the instability part, but the HF would have acceptable accuracy, especially at a distance of 40Mpc.

In this first study, we have considered a simple setup for the BNS simulations, where only hydrodynamics is taken into account and the effect of magnetic fields, bulk viscosity or other viscous effect are ignored. It will be important to perform detailed studies taking into account all physical effects, to determine which ones allow or prohibit the development of rotational instabilities. Future observational studies will either confirm the presence of rotational instabilities or set upper limits and constrain physical properties that contribute to their suppression.

\acknowledgements
We are grateful to Miquel Miravet for carefully reviewing this manuscript and for the fruitful conversation about \texttt{\textsc{BayesWave}} injections. We also thank Katerina Chatziioannou for useful comments and clarifications on the paper in Ref. \cite{Chatziioannou_2017} and Meg Millhouse, Sophie Hourihane for more information about the \texttt{\textsc{BayesWave}} pipeline. We acknowledge Teng Zhang, Huan Yang, Denis Martynov, Patricia Schmidt and Haixing Miao for providing us a \textsc{Python} notebook to reproduce the sensitivity curves in Ref.~\cite{HF}. AS acknowledges the Bodossaki Foundation for support in the form of a Ph.D. scholarship.   Figures in this manuscript were produced
using \texttt{Matplotlib}~\cite{plt}. The package \texttt{PyCBC}~\cite{pycbc} were used for benchmarking of different  waveform  models, for transformations between the frequency and time domain, and for other results. We are grateful for the computational resources provided by the LIGO Laboratory and supported by the U.S. National Science Foundation Awards PHY-0757058 and PHY-0823459. Virgo is funded, through the European Gravitational Observatory (EGO), by the French Centre National de Recherche Scientifique (CNRS), the Italian Istituto Nazionale di Fisica Nucleare (INFN) and the Dutch Nikhef, with contributions by institutions from Belgium, Germany, Greece, Hungary, Ireland, Japan, Monaco, Poland, Portugal, Spain. KAGRA is supported by Ministry of Education, Culture, Sports, Science and Technology (MEXT), Japan Society for the Promotion of Science (JSPS) in Japan; National Research Foundation (NRF) and Ministry of Science and ICT (MSIT) in Korea; Academia Sinica (AS) and National Science and Technology Council (NSTC) in Taiwan.
\appendix
\section{Injected waveform SNR for different cases \label{sec:appendixa}}
\begin{figure*}[h]
 	\includegraphics[width=.9\linewidth]{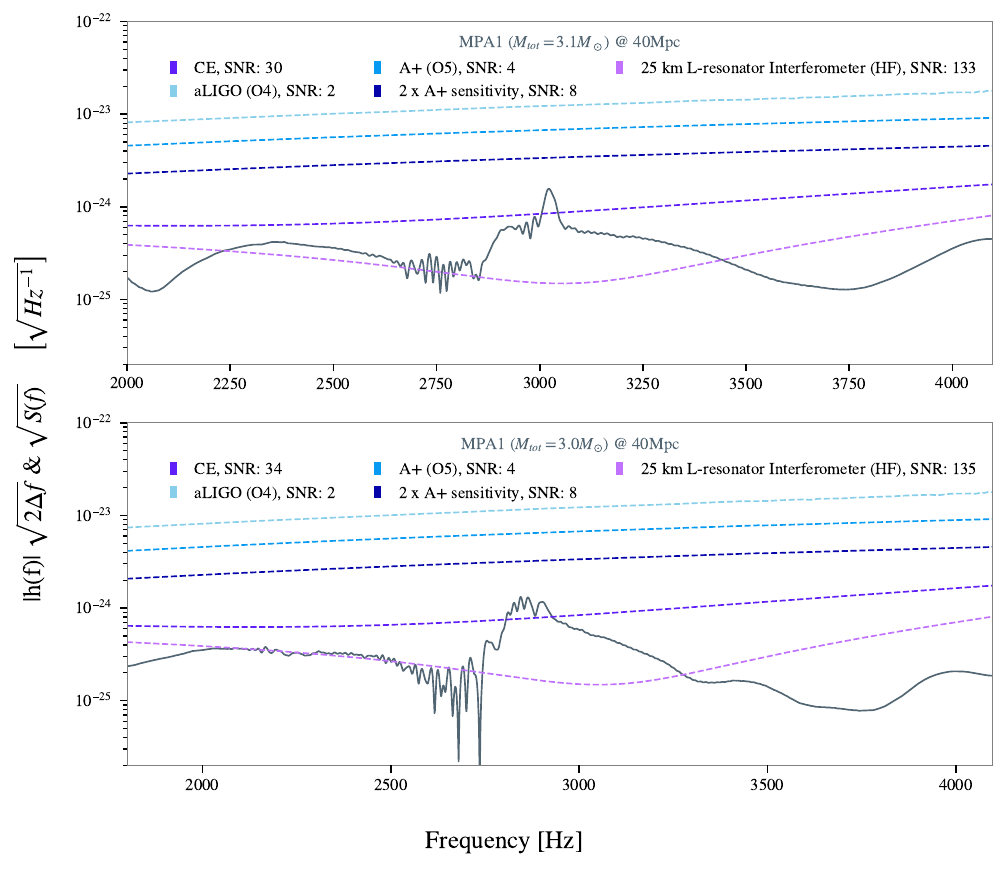}
 	\caption{Injected signal (see Table \ref{tab:injected_models}) in the frequency domain and the  sensitivity curve for each network configuration or detector design in Table \ref{tab:network_conf}. In each case, we mention the SNR of the entire post-merger phase. {\it Top panel}: $M_{\rm tot}=3.1M_\odot$ at 40Mpc. {\it Bottom panel}: $M_{\rm tot}=3.0M_\odot$ at 40Mpc. }
 \label{fig:inj_sens}
\end{figure*}
The spectrum of the waveforms (see Table \ref{tab:injected_models}) and the Amplitude Spectral Density (ASD) of each detector sensitivity (Table~\ref{tab:network_conf}), plus the sensitivity curves for O4 and O5) are presented in Fig.~\ref{fig:inj_sens}. The waveforms correspond to a BNS (MPA1 EoS) with a total mass of 3.1$M_\odot$ (top panel) and 3.0$M_\odot$ (bottom panel). The event is considered at 40 Mpc and is optimally oriented with respect to the detector. Labeled SNRs correspond to the optimal SNR of the {\it entire} post-merger phase.

For instance, the optimal post-merger SNR for the MPA1 EoS with $M_{\rm{tot}}$ = 3.1 \(M_\odot\) is computed via 
\begin{equation}
\left(\frac{S}{N}\right)^2=4 \int_{2000}^{4096} d f \frac{|\tilde{h}(f)|^2}{S_n(f)}
\label{eq:optimal_post_merger_snr_example_155}
\end{equation}
where the limits in the integral are in Hz.

From Fig. \ref{fig:inj_sens}, one can indicate that there is a poor possibility of obtaining valuable information about shear instability during O4 (SNR = 2) and O5 (SNR = 4). Only with a sensitivity twice the sensitivity of A + or better could an SNR of at least 8 be obtained, which is necessary for the detection of sigle events (see, e.g., ~\cite{Fiona_2023}). 

\section{Analysis for the full signal \label{sec:appendixb}}
\begin{figure}[t]
  \subfloat[Network configuration $2\times\rm{O}5$ sensitivity: L1 channel.]{%
\includegraphics[clip,width=0.95\columnwidth]{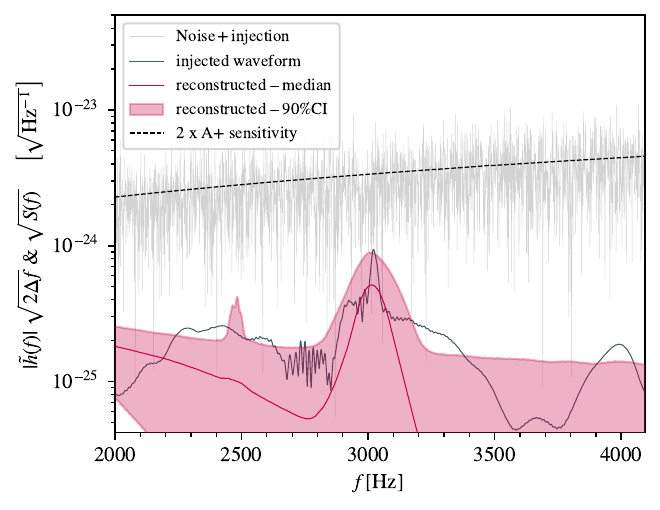}%
}

    \subfloat[Network configuration $\rm{CE+ET}$: CE channel.]{%
\includegraphics[clip,width=0.95\columnwidth]{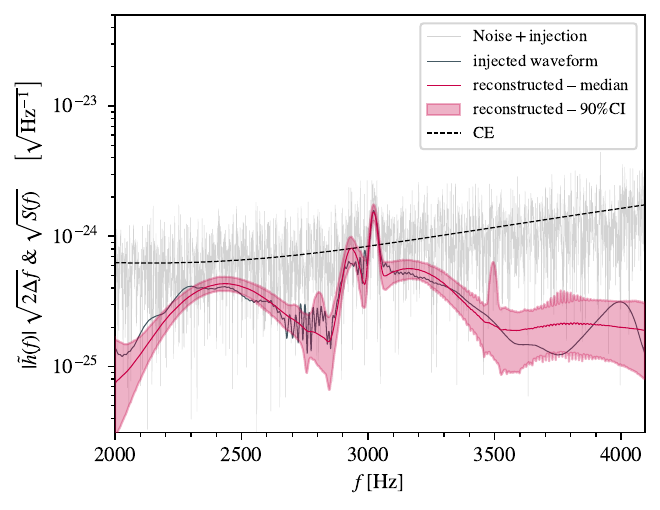}%
}

 	\caption{Reconstruction in the frequency domain of the entire post-merger phase for an optimally-oriented BNS merger with EoS MPA1 and $M_{\rm{tot}}=3.1~M_\odot$ at 40Mpc. {\it Top panel}: L1 channel for a network configuration with twice the sensitivity of the planned O5 run. {\it Bottom panel}: CE channel for the CE+ET network configuration.
}
 \label{fig:FD_40_MPA1_155155}
\end{figure}

\begin{figure}[t]
  \subfloat[Network configuration $2~\times~\rm{O}5$: L1 channel.]{%
\includegraphics[clip,width=0.95\columnwidth]{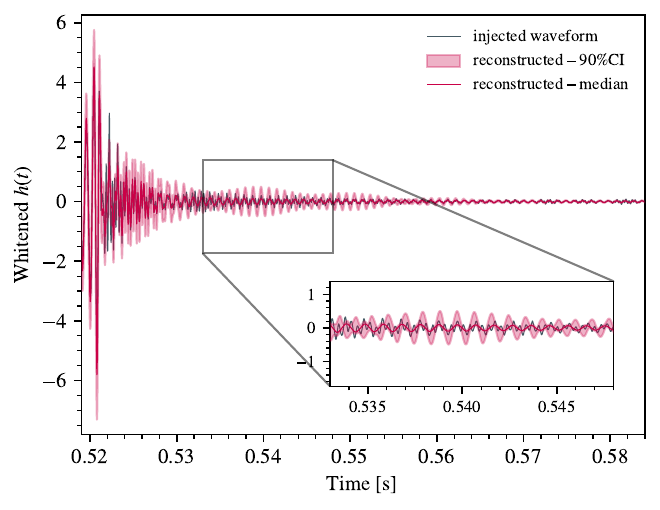}%
}

    \subfloat[Network configuration $\rm{CE+ET}$: CE channel.]{%
\includegraphics[clip,width=0.95\columnwidth]{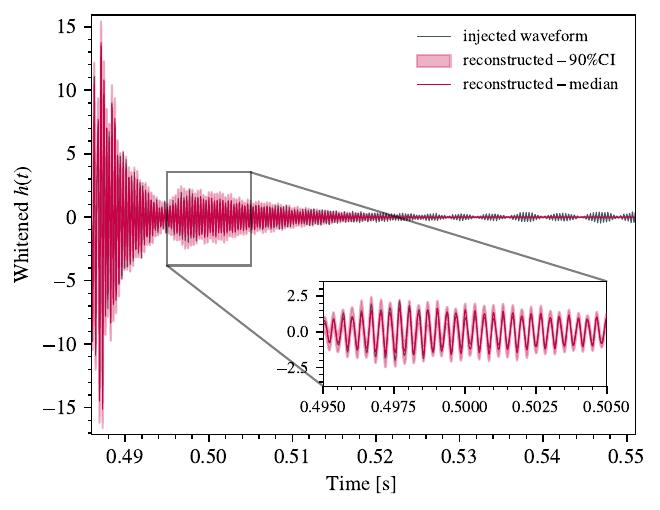}%
}

 	\caption{Same as Fig. \ref{fig:FD_40_MPA1_155155}, but in the time domain.}
 \label{fig:TD_40_MPA1_155155}
\end{figure}

In this Appendix, we consider the same analysis method mentioned in Sect.~\ref{sec:method}, but without cropping the signal and extracting only the instability part; \textit{instead, we perform the analysis in the entire post-merger phase (full injected signal)}. Figs.~\ref{fig:FD_40_MPA1_155155} and~\ref{fig:TD_40_MPA1_155155} represent the case of a BNS merger with the MPA1 EoS and $M_{\rm{tot}}=3.1~ M_\odot$ at 40 Mpc. The network cases correspond to (a) ``$2~\times~\rm{O}5$" and (b) ``CE+ET" (see Table~\ref{tab:network_conf}) and the figures refer to the L1 and CE channels, respectively. Figures show the injected waveform (gray), the median reconstructed waveform (pink), and its 90\% CI in the frequency domain (Fig.~\ref{fig:FD_40_MPA1_155155}) and time domain (Fig.~\ref{fig:TD_40_MPA1_155155}). Additionally, in Fig.~\ref{fig:FD_40_MPA1_155155}, the analyzed data (light gray), which is labeled as ``Noise+injection", are shown. The noise is Gaussian and is generated with respect to the ASD sensitivity (black dashed line). Both figures indicate the capability of all detectors to recover the post-merger GW signal. 

For the case of the ``$2~\times~\rm{O}5$" network configuration, Fig.~\ref{fig:FD_40_MPA1_155155} reveals the capacity of the network to capture the peak frequency, while the "early" post-merger is well reconstructed. The performance of the ``$\rm{CE+ET}$" network seems very good.

For a quantitative evaluation, we calculate the overlap $\mathcal{O}$ between the entire injected signal and the median reconstructed signal as a function of the optimal post-merger SNR. The overlaps corresponding to the ``$2~\times~\rm{O}5$" LIGO detector are good, despite the fact that we note difficulties in recovering the shear instability. The overlap is larger due to the good reconstruction of the early post-merger signal. In both mass models, the CE detector reaches ${\rm SNR}\sim 48$ with an overlap of ${\cal O} \simeq 0.99$, while the HF design reaches ${\rm SNR}\sim 135$ with an overlap of ${\cal O} \simeq 0.999$.
\begin{figure}[t]
  \subfloat[$M_{\rm{tot}}=3.0$ $M_\odot$.]{%
\includegraphics[clip,width=0.67\columnwidth]{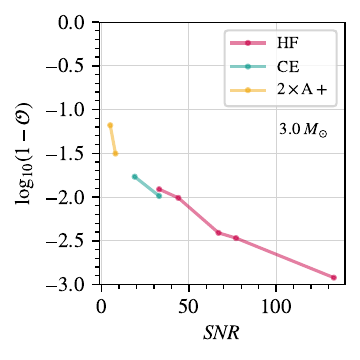}%
}

    \subfloat[$M_{\rm{tot}}=3.1$ $M_\odot$.]{%
\includegraphics[clip,width=0.67\columnwidth]{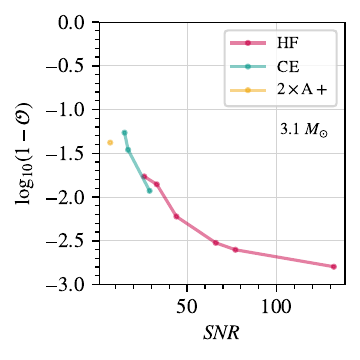}%
}
 	\caption{Overlap of the median reconstructed signal for the entire post-merger phase for a BNS merger with MPA1 EoS and for different detector networks or designs. 
  {(a)} $M_{\rm{tot}}=3.0$ $M_\odot$ ({\it top panel}),  {(b)} $M_{\rm{tot}}=3.1$ $M_\odot$ ({\it bottom panel}).}
 \label{fig:overlap_full_signal}
\end{figure}

\begin{figure}[htp]
  \subfloat[$M_{\rm{tot}}=3.0$ $M_\odot$.]{%
\includegraphics[clip,width=1\columnwidth]{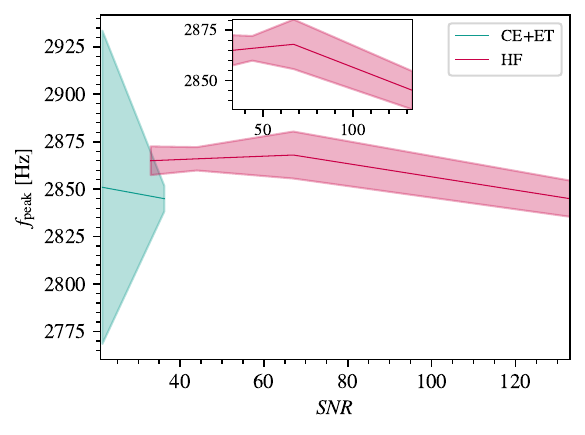}%
}

    \subfloat[$M_{\rm{tot}}=3.1$ $M_\odot$.]{%
\includegraphics[clip,width=1\columnwidth]{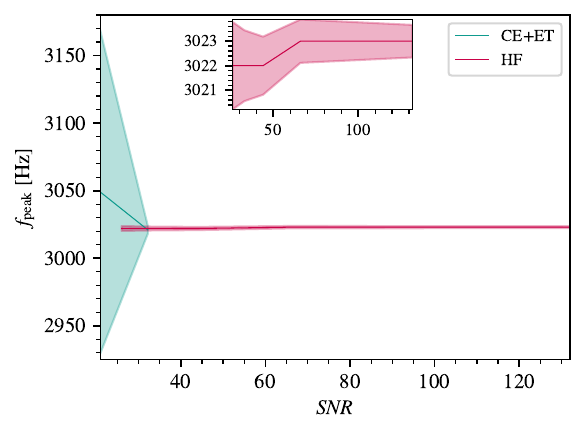}%
}
 	\caption{Median recovered frequency peak ($f_{\rm{peak}}$) using the entire post-merger phase for for \textbf{(a)} $M_{\rm{tot}}=3.0$ $M_\odot$ and  \textbf{(b)} $M_{\rm{tot}}=3.1$ $M_\odot$. The shaded area refers to the standard deviation $\pm$. Two different detector cases are compared.}
 \label{fig:fpeak}
\end{figure}

The median recovered peak frequency $f_{\rm{peak}}$ (in Hz) for the MPA1 EoS and for both mass cases, as a function of the optimal post-merger SNR is shown in Fig.~\ref{fig:fpeak}. For the case of a BNS merger with a total mass of 3.0 \(M_\odot\) (Fig. \ref{fig:fpeak}a),  the recovered $f_{\rm{peak}}$ reaches an uncertainty of $\Delta f_{\rm{peak}}\sim 35$ Hz for the HF design. The ``CE+ET" network configuration seems to be able to give similar results for an optimal post-merger network SNR 37. For lower SNRs, the data become more uninformative and the error in the peak frequency becomes higher.

When the total mass of the BNS is 3.1 \(M_\odot\) (Fig. \ref{fig:fpeak} b), the HF detector captures almost perfectly the peak frequency ($\Delta f_{\rm{peak}}\lesssim 3.4$Hz), resulting in good estimates up to 200 Mpc (post-merger SNR $\sim 34$). Regarding the performance of the ``CE+ET" network, the peak frequency is well estimated when the source is optimally oriented at 40 Mpc (post-merger network SNR $\sim37$).

\clearpage
%\bibliographystyle{apsrev4-1}
%\bibliography{bib}

\end{document}